\documentclass[twocolumn,aps,preprintnumbers,amsmath,amssymb]{revtex4}

\usepackage{graphicx}
\usepackage{dcolumn}
\usepackage{bm}
\usepackage{color}

\begin{document}


\title{Observation of a new field-induced phase transition and its concomitant quantum critical fluctuations in CeCo(In$_{\bm{1-x}}$Zn$_{\bm{x}}$)$_{\bm{5}}$}
\author{Makoto~Yokoyama}
\email[]{makoto.yokoyama.sci@vc.ibaraki.ac.jp}
\author{Hiroaki~Mashiko}
\author{Ryo~Otaka}
\author{Yoshiki~Oshima}
\author{Kohei~Suzuki}
\affiliation{Faculty of Science, Ibaraki University, Mito, Ibaraki 310-8512, Japan}
\author{Kenichi~Tenya}
\affiliation{Faculty of Education, Shinshu University, Nagano 380-8544, Japan}
\author{Yusei~Shimizu}
\author{Ai~Nakamura}
\author{Dai~Aoki}
\affiliation{Institute for Materials Research, Tohoku University, Oarai, Ibaraki 311-1313, Japan}
\author{Akihiro~Kondo}
\author{Koichi~Kindo}
\author{Shota~Nakamura}
\author{Toshiro~Sakakibara}
\affiliation{Institute for Solid State Physics, The University of Tokyo, Kashiwa, Chiba 277-8581, Japan}

\date{\today}
             
\begin{abstract}
We demonstrate a close connection between observed field-induced antiferromagnetic (AFM) order and quantum critical fluctuation (QCF) in the Zn7\%-doped heavy-fermion superconductor CeCoIn$_5$. Magnetization, specific heat, and electrical resistivity at low temperatures all show the presence of new field-induced AFM order under the magnetic field $B$ of 5--10 T, whose order parameter is clearly distinguished from the low-field AFM phase observed for $B < 5\ {\rm T}$ and the superconducting phase for $B < 3\ {\rm T}$. The 4f electronic specific heat divided by the temperature, $C_e/T$, exhibits $-\ln T$ dependence at $B\sim 10\ {\rm T}$  ($\equiv B_0$), and furthermore, the $C_e/T$ data for $B \ge B_0$ are well scaled by the logarithmic function of $B$ and $T$:  $\ln[(B-B_0)/T^{2.7}]$. These features are quite similar to the scaling behavior found in pure CeCoIn$_5$, strongly suggesting that the field-induced QCF in pure CeCoIn$_5$ originates from the hidden AFM order parameter equivalent to high-field AFM order in Zn7\%-doped CeCoIn$_5$.
\end{abstract}

\maketitle

\section{Introduction}
Understanding the role of quantum critical fluctuations (QCFs) in unconventional superconductivity is a longstanding subject in condensed matter physics. Superconducting (SC) order commonly emerges near the magnetic phases in most of the strongly correlated electron systems, such as high-$T_c$ cuprates \cite{rf:Lee2006}, FeAs-based alloys \cite{rf:Stewart2011}, and heavy-fermion compounds \cite{rf:Mathur98,rf:Pfleiderer2009}. Thus, it is considered that QCFs arising from magnetic correlation play a crucial role in the Cooper pairing in these compounds. Heavy fermion superconductors are particularly suitable for research on the interplay between QCFs and SC order, because magnetic and SC orders can be easily tuned by the magnetic field and the pressure, as well as chemical substitutions, owing to the small energy scales of these orders.

The ternary tetragonal compound CeCoIn$_5$ is one of the most extensively investigated heavy fermion superconductor ($T_{sc}=2.3\ {\rm K}$) from the perspective of the interplay between antiferromagnetic (AFM) QCF and SC order \cite{rf:Petrovic2001}. This compound shows non-Fermi-liquid (NFL) behaviors in various quantities just above the SC upper critical field $\mu_0H_{c2}$ ($=5\ {\rm T}$), such as nearly $T$-linear dependence in magnetization and electrical resistivity \cite{rf:Tayama2002,rf:Paglione2003}, and $-\ln T$ divergence in the specific heat divided by the temperature \cite{rf:Bianchi2003-2}, by applying magnetic field $B$ along the $c$ axis. It is widely recognized that AFM-QCF enhanced near $H_{c2}$ is responsible for these NFL behaviors. However, relevant AFM order is never found in CeCoIn$_5$, except for SC order coupled with AFM spin modulation, called the Q phase, located in small temperature and magnetic field ranges just below $H_{c2}$ for $B\perp c$ \cite{rf:Kakuyanagi2005,rf:Young2007,rf:Kenzelmann2008}.

Instead, long-range AFM orders with various spin modulations are actually induced by substituting ions in the elements of CeCoIn$_5$, such as Nd for Ce \cite{rf:Hu2008,rf:Raymond2014}, Rh for Co \cite{rf:Zapf2001,rf:Yoko2006,rf:Ohira-Kawamura2007,rf:Yoko2008}, and Cd, Hg, and Zn for In \cite{rf:Pham2006,rf:Nicklas2007,rf:Yoko2014}, although it is still unclear how they are related to the QCF observed around $H_{c2}$ in CeCoIn$_5$ \cite{rf:Seo2014,rf:Nair2010}. We reported that the mixed compound CeCo(In$_{0.93}$Zn$_{0.07}$)$_5$ shows AFM order below $T_N\sim 2.2$ K, along with the reduction of $T_{sc}$ down to 1.3 K \cite{rf:Yoko2014,rf:Yoko2015}. In a previous study, however, the issue of the relationship between QCF and AFM order was left unresolved. In particular, the specific heat divided by the temperature for Zn7\%-doped CeCoIn$_5$ shows $-\ln T$ dependence at $B\sim$ 10 T for $B\,||\,c$, whose temperature dependence coincides fairly well with that observed around $H_{c2}$ in pure CeCoIn$_5$ \cite{rf:Yoko2015}. The origin of this NFL anomaly has not yet been clarified, because the applied field range for yielding this NFL anomaly ($B\sim$ 10 T) is much larger than the critical fields of AFM order (5 T) and SC order (3 T).

In this paper, we report the finding of another new AFM order in proximity to the field-induced NFL anomaly in Zn7\%-doped CeCoIn$_5$, which is clearly distinguished from the low-field AFM order previously observed for $B \le 5\ {\rm T}$, as revealed by thermodynamic evidence. This is the first manifestation of the close connection between field-induced  QCF and AFM order in CeCoIn$_5$ and its substituted systems. Furthermore, the scaling analysis for specific heat demonstrates that the NFL behavior around the observed AFM order is quite similar to that seen around $H_{c2}(T\to 0)$ in CeCoIn$_5$. This resemblance strongly suggests that field-induced QCFs in pure and Zn-doped CeCoIn$_5$ have almost the same origin.

\section{Experiment Details}
Single crystals of CeCo(In$_{1-x}$Zn$_x$)$_5$ with $x=0.07$ were grown with the indium-flux technique, whose details are described elsewhere \cite{rf:Yoko2015}. Note that the actual Zn concentration, $y$, estimated from the energy dispersive x-ray spectroscopy (EDS) measurements was roughly $0.025$, although it involves the fairly large uncertainty unavoidable in EDS measurements \cite{rf:Yoko2015}. Here, we use the nominal concentration $x$ for clarity and simplicity. The magnetization $M$ measurement was performed down to 0.08 K using a capacitively-detected Faraday force magnetometer \cite{rf:Sakakibara94}. The $a$-axis electrical resistivity $\rho$ was measured in the temperature range of 0.06--0.26 K, using the standard four-wire technique. The specific heat $C_p$ measurement was carried out down to 0.5 K with the thermal relaxation method using a commercial measurement system (PPMS, Quantum Design). In all measurements, magnetic field $B$ ($\mu_0H$) was applied up to 14 T along the $c$ axis.  

\section{Results and Discussion}
Figure 1(a) shows the $c$-axis magnetization curve, $M(B)$, at 0.08 K, measured under increasing and decreasing field processes. The SC upper critical field and the subsequent change in the AFM structure are recognized by the closing of the large hysteresis loop at $\mu_0H_{c2} =3\ {\rm T}$ and a step-like anomaly at $B_{M1}=5\ {\rm T}$ in $M(B)$, respectively \cite{rf:Yoko2015}. The present $M(B)$ measurement further finds a weak kink or bending at $B_{M2}=9.8(6)\ {\rm T}$. This anomaly is also confirmed by the broad step-like decrease in the field variation of $dM/dB$ [the inset of Fig.\ 1(a)]. 

To clarify the origin of the kink at $B_{M2}$ in $M(B)$, we examine the temperature variations of the specific heat, $C_p(T)$, for $B \ge B_{M1}$. As shown in Fig.\ 1(b), $C_p(T)$ for $B=7$ T exhibits a jump-like anomaly at $T_{M2}\sim 0.9\ {\rm K}$, evidencing the emergence of an ordered phase below $T_{M2}$ for $B > B_{M1}$. The jump in $C_p$ moves toward a lower temperature with increasing $B$, and its magnitude becomes small at $B=8\ {\rm T}$. The jump finally disappears at least for $T> 0.5\ {\rm K}$ at $B=9\ {\rm T}$ ($\sim B_{M2}$). Because $T_{M2}$ is reduced toward zero as $B$ approaches $B_{M2}$, the kink seen at $B_{M2}$ in $M(B)$ is considered to be ascribed to the breakdown of the same ordered phase.

The development of the ordered phase for $B \ge B_{M1}$ also yields a broad peak at $T_{M2}=0.9\ {\rm K}$ ($B=7\ {\rm T}$) and 0.6 K (8 T) in the temperature variations of $M/B$ [Fig.\ 1(c)]. Because the magnitude of $M/B$ is reduced below $T_{M2}$, it is natural to consider that the ordered phase for $B \ge B_{M1}$ is attributed to the AFM ordering. For $B > B_{M2}$, however, $M/B$ follows the $-T^n$ function, and the best fit in the temperature range below 2.2 K gives the exponent $n$ of 1.38(1) at $B=10\ {\rm T}$. The $n$ value increases with increasing $B$ and becomes 1.58(1) at $B=14\ {\rm T}$.
\begin{figure}[tbp]
\begin{center}
\includegraphics[bb=48 170 411 757,keepaspectratio,width=0.45\textwidth]{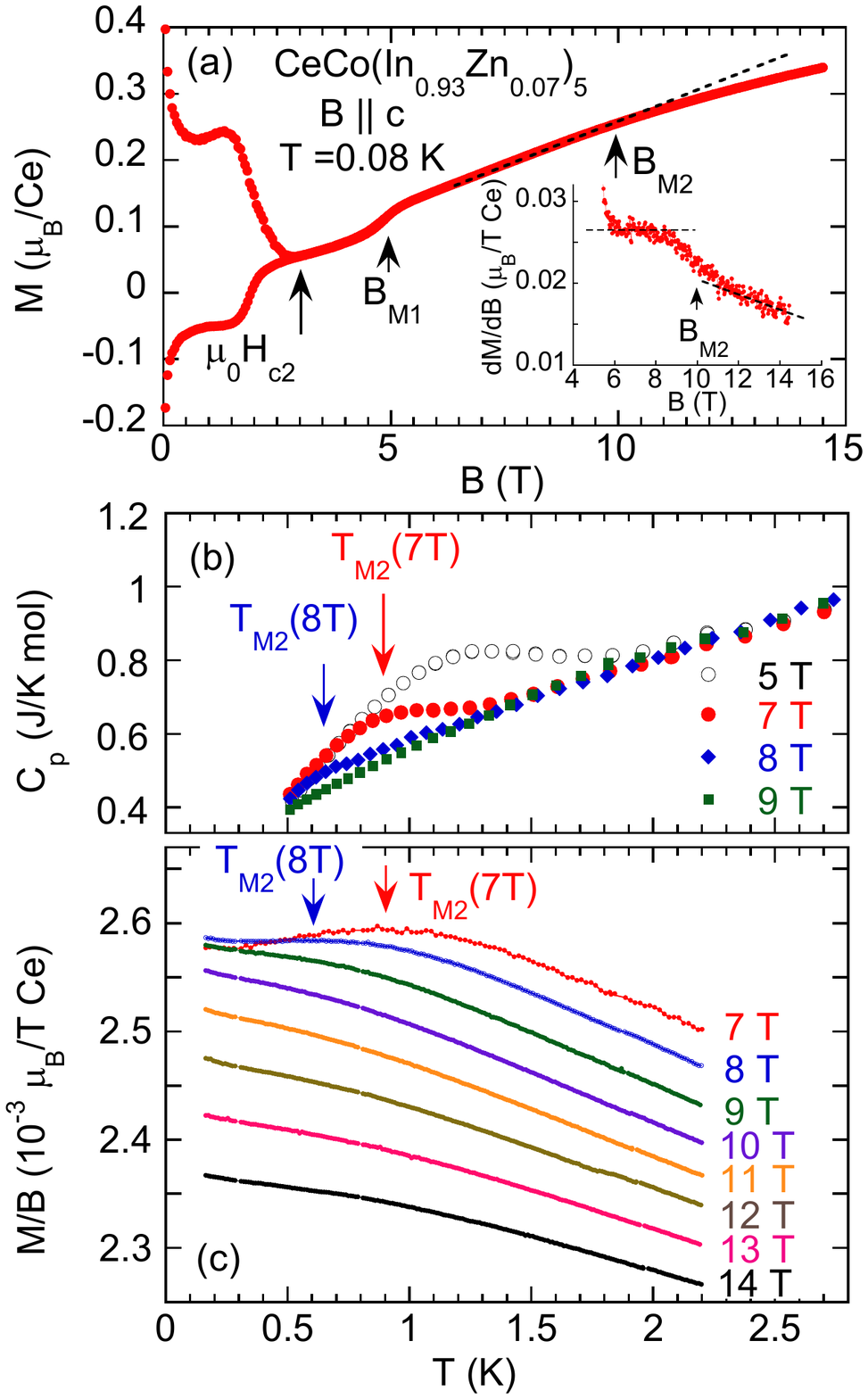}
\end{center}
  \caption{
(Color online)  (a) Magnetic field dependence of the $c$-axis magnetization $M$ at 0.08 K for CeCo(In$_{0.93}$Zn$_{0.07}$)$_5$. The inset of (a) shows the $dM/dB$ plot around 10 T. The broken lines are guides for the eye. The temperature variations of (b) the specific heat, $C_p$, for $B\,||\,c$, and (c) the $c$-axis magnetization divided by the magnetic field, $M/B$, for CeCo(In$_{0.93}$Zn$_{0.07}$)$_5$.
}
\end{figure}

\begin{figure}[tbp]
\begin{center}
\includegraphics[bb=50 394 484 726,keepaspectratio,width=0.45\textwidth]{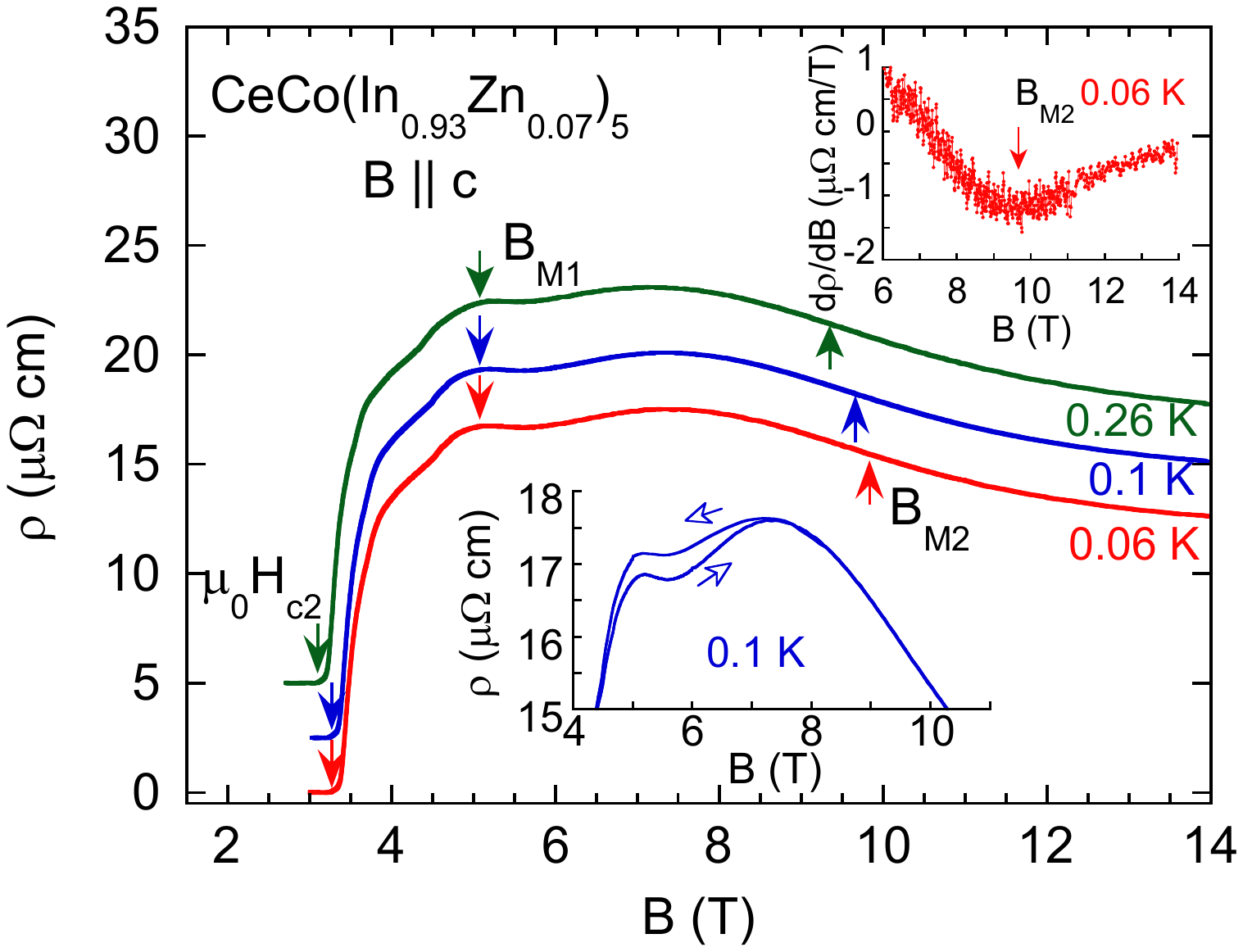}
\end{center}
  \caption{
(Color online)  Field variations of the $a$-axis electrical resistivity $\rho(B)$ for CeCo(In$_{0.93}$Zn$_{0.07}$)$_5$, obtained with the increasing $B$ sweep for $B\,||\,c$. Here, the vertical baselines of the $\rho$ data for $T\ge 0.1\ {\rm K}$ are shifted in steps of 2.5 $\mu\Omega$ cm for clarity. The upper inset shows $d\rho/dB$ at 0.06 K, and the lower inset is the $\rho$ data at 0.1 K, taken under the increasing and decreasing $B$ sweeps. The arrows in the lower inset indicate the directions of the $B$ sweep. 
}
\end{figure}
The magnetic field dependences of the electrical resistivity $\rho(B)$ for $B\,||\,c$ and $T\le 0.26\ {\rm K}$ are plotted in Fig.\ 2. The $\rho(B)$ curve has a shoulder-like structure at $B_{M1}$ and an inflection point at $\sim B_{M2}$. The latter anomaly is confirmed by the minimum in $d\rho/dB$ (the upper inset of Fig.\ 2). The inflection point in $\rho(B)$ moves from 9.8(3) T ($T=0.06\ {\rm K}$) to 9.3(3) T ($T=0.26\ {\rm K}$) with increasing temperature, and this temperature variation is consistent with the tendency in the field variations of $T_{M2}$ derived from $C_p$ and $M/B$. $\rho(B)$ is found to show hysteretic behavior (the lower inset of Fig.\ 2). Hysteresis appears at $\sim B_{M1}$ in $\rho(B)$ and then gradually becomes small with increasing $B$. At 0.1 K, hysteresis is still observed at least up to 7.7 T but cannot be detected at $B_{M2}$ within the experimental resolution.  The absence of the hysteresis at $B_{M2}$ in $\rho(B)$ implies the second-order nature of the phase transition at $B_{M2}$. The continuous changes in $M(B)$ and $\rho(B)$ at $B_{M2}$ are also characteristic of the second-order transition.

In Fig.\ 3, we summarize the magnetic field versus temperature ($B-T$) phase diagram obtained from the specific heat, electrical resistivity, and dc magnetization measurements. Large parts of the boundaries of low-field AFM and SC phases were clarified in a previous study \cite{rf:Yoko2015}. The present investigation reveals, for the first time, the emergence of another high-field AFM phase located between $B_{M1}$ and $B_{M2}$. Its boundary with the paramagnetic (PM) state, $B_{M2}(T)$, decreases with the increasing temperature and seems to meet the boundary between the low-field AFM and PM states around $T=1.5\ {\rm K}$ and $B=4.5\ {\rm T}$. The phase transition at $B_{M1}$, observed as the step-like anomaly in $M(B)$, separates the high-field AFM phase from the low-field AFM phase. In addition, the shapes of the high-field and low-field AFM regions in the $B-T$ phase diagram are qualitatively different from each other. These two features suggest a discrepancy in the AFM spin structures between these AFM orders.
\begin{figure}[tbp]
\begin{center}
\includegraphics[bb=34 468 458 798,keepaspectratio,width=0.45\textwidth]{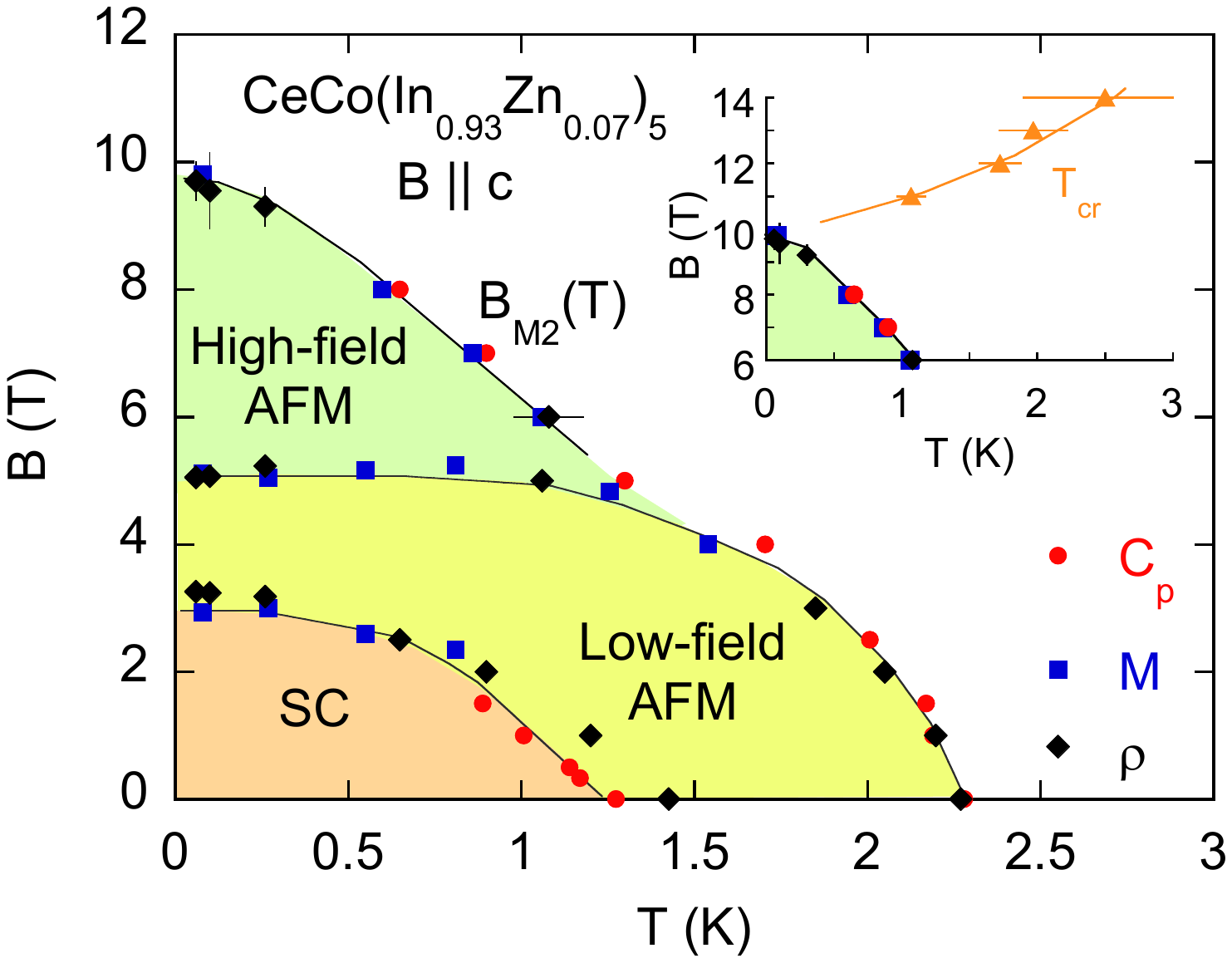}
\end{center}
  \caption{
(Color online)  $B-T$ phase diagram of CeCo(In$_{0.93}$Zn$_{0.07}$)$_5$ for $B\,||\,c$, derived from the specific heat, electrical resistivity, and dc magnetization measurements \cite{rf:Yoko2015}. The inset shows the enlargement of the $B-T$ phase diagram around 10 T. The lines on the phase boundaries are guides for the eye. 
}
\end{figure}

We now focus on the relationship between the observed high-field AFM order and the NFL behavior seen around $B_{M2}$ in the specific heat, by performing precise $C_p$ measurements and analyses. Figure 4(a) shows the temperature dependence of the 4f electronic specific heat divided by the temperature, $C_e/T$. In these data, the phonon and nuclear-spin contributions are carefully subtracted using $C_p$ data of the non-4f compound LaCoIn$_5$ at $B=0$ and calculations based on the natural abundance of nuclear spins, respectively. Note that the phonon contribution in $C_p$ is about 10\% even at 4 K, and therefore, the effect of Zn doping ($y\sim 0.025$) is considered to be negligible at low temperatures. However, we do not estimate $C_e$ below 0.5 K because the nuclear spin contribution is expected to be very large and become more than 95\% of $C_p$ at 0.1 K for $B \ge 10\ {\rm T}$. $C_e/T$ shows clear $-\ln T$ dependence at 10 T ($\simeq B_{M2}$) below 4 K, and this $T$ dependence is gradually suppressed for $B> 10\ {\rm T}$. This suppression is considered to be a crossover from the NFL to Fermi-liquid (FL) states for $B > B_{M2}$. The appearance of the NFL-FL crossover may also yield the increase in the exponent $n$ in $M/B \propto -T^n$. We define the crossover temperature $T_{cr}$ as the onset of the deviation from the $-\ln T$ function in $C_e/T$. As seen in the inset of Fig.\ 3, the curve corresponding to $T_{cr}$ approaches the high-field AFM boundary for $T\to 0$. This feature suggests that the QCF responsible for the NFL behavior in $C_e/T$ still exists at $T\sim 0$ and $B\sim B_{M2}$, and thus, the phase boundary of the high-field AFM phase at $T=0$ and $B=B_{M2}$ is regarded as the quantum critical point.
\begin{figure}[tbp]
\begin{center}
\includegraphics[bb=11 318 552 706,keepaspectratio,width=0.5\textwidth]{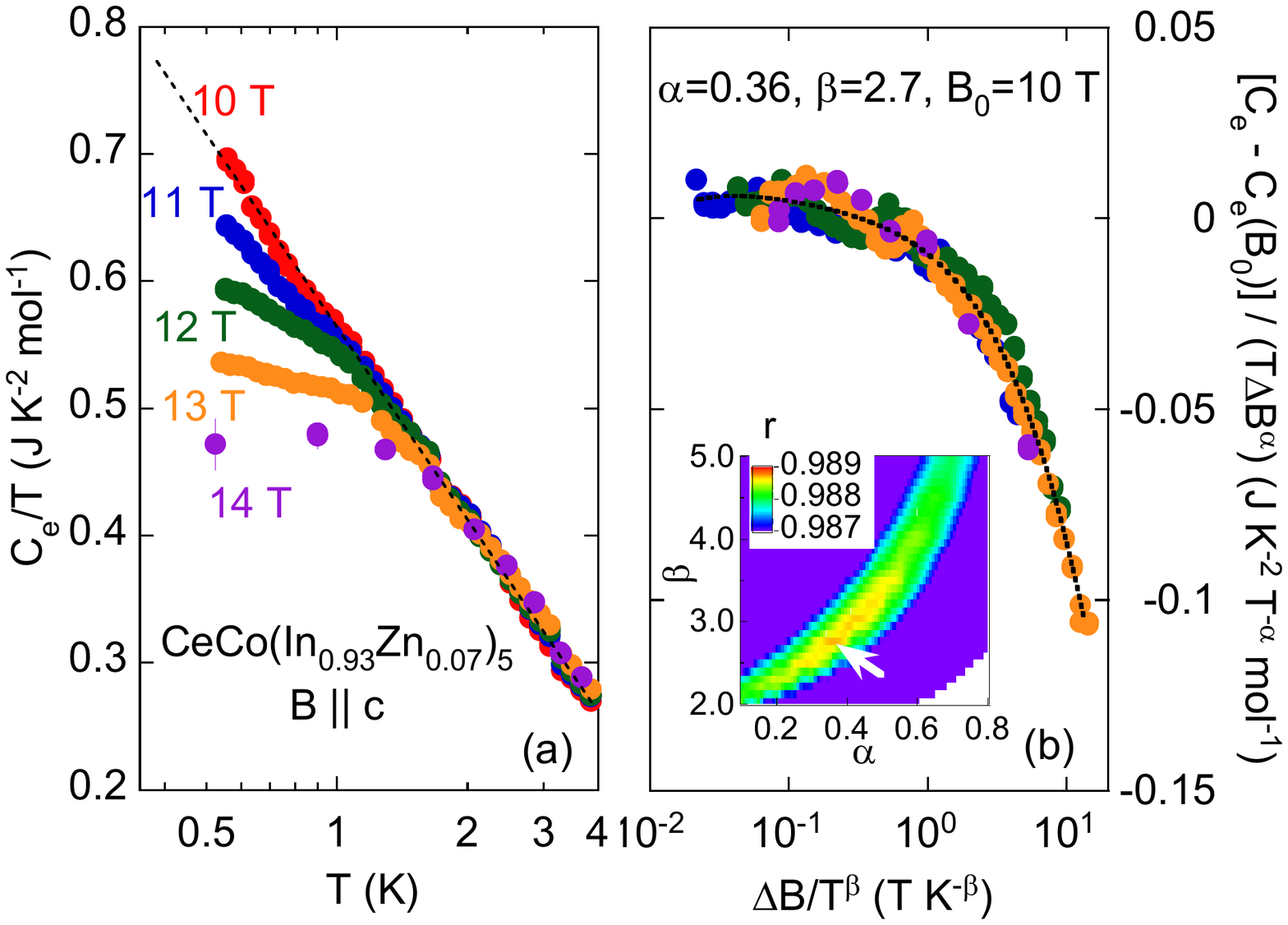}
\end{center}
  \caption{
(Color online)  (a) Temperature variations of the 4f electronic specific heat divided by the temperature, $C_e/T$, for CeCo(In$_{0.93}$Zn$_{0.07}$)$_5$, obtained under magnetic fields above 10 T for $B\,||\,c$. The broken line indicates a guide for the $-\ln T$ function. (b) The plot of $[C_e-C_e(B_0)]/(T \Delta B^\alpha)$ versus $\Delta B/T^\beta$ obtained with scaling analysis of the $C_e/T$ data, where $\Delta B$ indicates $B-B_0$ with $B_0=10\ {\rm T}$. The inset of (b) is the image plot of the correlation coefficient $r$, resulting from the least squares fitting of the $C_e/T$ data for the scaling analysis. The position of the maximum $r$  ($\alpha=0.36$ and $\beta=2.7$) is given by the arrow, and the fitting curve with these $\alpha$ and $\beta$ values is shown as the broken line in (b).
}
\end{figure}

It is interesting to perform scaling analysis for the $C_e/T$ data in order to find similarities and differences in field-induced QCFs between pure and Zn7\%-doped CeCoIn$_5$. Here, the scaling function with the form of $[C_e-C_e(B_0)]/(T\Delta B^\alpha)= f[\ln(\Delta B/T^\beta)]$ with $\Delta B=B-B_0$ is assumed in accordance with the procedure executed for pure CeCoIn$_5$ \cite{rf:Bianchi2003-2}. We set $B_0$ to 10 T and tentatively use a fourth-order polynomial function for $f[\ln(\Delta B/T^\beta)]$ to perform the curve fitting with least squares. The correlation coefficient of the fits, $r$, for the given scaling parameters $\alpha$ and $\beta$ are plotted in the inset of Fig.\ 4(b). In Fig.\ 4(b), we show the best scaling result with the scaling parameters $\alpha=0.36$ and $\beta=2.7$, which are determined so that $r$ becomes the maximum. It covers the NFL and nearly FL features in the $C_e/T$ data between 10 and 14 T. The obtained $\beta$ value is comparable to the reported value ($\beta=2.5$) for pure CeCoIn$_5$, whereas $\alpha$ is significantly smaller than the reported value ($\alpha=0.71$). The discrepancy in the $\alpha$ value would simply arise from the large difference in $B_0$ between the Zn7\%-doped alloy ($B_0=10\ {\rm T}$) and the pure compound ($B_0=5\ {\rm T}$), since $\alpha$ concerns $\Delta B$ as its exponent and is not included in $ f[\ln(\Delta B/T^\beta)]$. However, the correspondence in $\beta$ leads to a quite interesting result; $C_e/T$ in pure and Zn7\%-doped CeCoIn$_5$ obey the same variable, $\ln(\Delta B/T^\beta)$, for $B \ge B_0$.

This scaling analysis strongly suggests that field-induced QCFs in pure and Zn7\%-doped CeCoIn$_5$ have almost the same origin. Furthermore, the present $C_p$ study of the latter compound reveals that the QCF appears around the quantum critical point of the high-field AFM phase. We thus consider that the QCF enhanced at $\sim H_{c2}$ in pure CeCoIn$_5$ also originates from the ``hidden" order parameter of the high-field AFM phase in CeCo(In$_{0.93}$Zn$_{0.07}$)$_5$.

Thus far, two possible scenarios concerning the quantum critical point have been suggested in pure CeCoIn$_5$. The scaling analysis for the magnetic Gr\"uneisen parameter suggests that the quantum critical point exists at $B\sim 0$ \cite{rf:Tokiwa2013}, although the $C_p$ and $\rho$ measurements propose that it is located at $\sim H_{c2}$ \cite{rf:Paglione2003,rf:Bianchi2003-2}. In this regard, the present study shows common characteristics of field-induced QCFs between pure and Zn7\%-doped compounds and thus supports the latter suggestion. However, the QCF enhanced at $\sim B_{M2}$ is not coupled with low-field AFM order in the Zn7\%-doped alloy, because this order is broken by the phase transition at $B_{M1}$ $(\ll B_{M2})$. We observed no indication of a QCF related to low-field AFM order in the Zn7\%-doped alloy. Instead, the QCF may be hidden deep inside the SC phase ($B \ll \mu_0H_{c2}$) in pure CeCoIn$_5$, because the $T_N$ of the low-field AFM transition decreases with the decreasing Zn concentration $x$ and seems to approach zero at $x\sim 0$ \cite{rf:Yoko2015}. This trend is also observed in CeCo(In,Cd)$_5$ \cite{rf:Pham2006}. 

The emergence of AFM order, coupled with the field-induced QCF, in the Zn7\%-doped alloy is in stark contrast to the absence of AFM order in the pure compound. We expect that this finding paves the way for microscopic research on the relationship between the AFM correlation and SC order. Research on the spin structures and excitations in AFM order of the Zn7\%-doped alloy is the key to understanding the microscopic nature of QCFs in pure and Zn-doped CeCoIn$_5$. At the same time, such research is expected to provide a clue for resolving the unusual SC properties related to the AFM spin correlations in CeCoIn$_5$, such as the Q phase \cite{rf:Kakuyanagi2005,rf:Young2007,rf:Kenzelmann2008} and spin resonance excitation \cite{rf:Stock2008,rf:Eremin2008,rf:Raymond2015,rf:Song2016,rf:Chubukov2008}. 

A possible spin structures of high-field AFM order in CeCo(In$_{1-x}$Zn$_x$)$_5$ is the incommensurate spin modulation identical to that seen in the Q phase. In this case, the present experimental results suggest that the spin arrangement of the Q phase does not have to coexist with superconductivity in Zn-doped CeCoIn$_5$, although the Q phase always coexists with the SC state in pure CeCoIn$_5$. Alternatively, it is also probable that high-field AFM order in CeCo(In$_{1-x}$Zn$_x$)$_5$ has a spiral spin structure similar to that seen in isostructural CeRhIn$_5$ \cite{rf:Bao2000}, or is composed of canted spins with the same wave vector as that of low-field AFM order. Elastic neutron scattering and nuclear magnetic resonance experiments will be the key to determining the high-field AFM structure. In addition, it would be interesting to investigate spin excitations in low- and high-field AFM phases for CeCo(In$_{1-x}$Zn$_x$)$_5$ using the inelastic neutron scattering technique. In pure CeCoIn$_5$, it was previously suggested that the spin resonance peak appears in connection with SC gap symmetry \cite{rf:Stock2008,rf:Eremin2008}. However, subsequent studies proposed that this peak originates from the instability of commensurate AFM order or the Q phase \cite{rf:Raymond2015,rf:Song2016,rf:Chubukov2008}. Comparing the characteristics of spin excitations between pure and Zn-doped CeCoIn$_5$ will provide a clue for resolving the issue of the spin resonance peak. In addition, measuring quantum oscillations and photoemission for CeCo(In$_{1-x}$Zn$_x$)$_5$ may elucidate how the occurrence of two AFM orders is coupled with the change in the electronic structure.

\section{Conclusion}
In conclusion, our $M$, $C_p$, and $\rho$ measurements for CeCo(In$_{0.93}$Zn$_{0.07}$)$_5$  revealed the emergence of field-induced AFM order between $B_{M1}=5\ {\rm T}$ and $B_{M2}=9.8\ {\rm T}$ for $B\,||\,c$. The specific heat exhibits characteristic NFL behavior ($C_e/T\propto -\ln T$) around $B_{M2}$, and then the FL state recovers with further increasing $B$. This variation in $C_e/T$ can be well scaled by the variable, $\ln(\Delta B/T^\beta)$, which is quite similar to the scaling behavior found in pure CeCoIn$_5$. This correspondence strongly suggests that the field-induced QCF near $H_{c2}$ in pure CeCoIn$_5$ comes from the instability of the hidden AFM order parameter, which emerges as high-field AFM order in Zn7\%-doped CeCoIn$_5$. To clarify the connection of the QCFs between pure and Zn-doped CeCoIn$_5$ more precisely, we plan to investigate the doping dependence of high-field AFM order and its fluctuations in CeCo(In$_{1-x}$Zn$_{x}$)$_5$ for $x<0.07$. 

\begin{acknowledgments}
M.Y. is grateful to I. Kawasaki, Y. Homma, and S. Kittaka for their experimental support. This study was supported in part by Grants-in-Aid for Scientific Research on Innovative Areas ``J-Physics" (15H05883) from MEXT and KAKENHI (15H03682 and 17K05529) from JSPS.
\end{acknowledgments}

\end{document}